\documentclass{article}

\usepackage{graphicx,color,wrapfig}
\usepackage{amssymb,amsmath}
\usepackage{url}
\usepackage{xspace}
\usepackage{float}
\usepackage{graphics}
\usepackage{soul}
\usepackage{subcaption}
\usepackage{textcomp} 

\chardef\other=12

\def\mmakeactive#1{\catcode`#1=\active\ignorespaces}

{
\mmakeactive\
\gdef\obeywhitespace{%
  \mmakeactive\^^M %
  \let^^M=\NewLine %
  \aftergroup\removebox %
  \obeyspaces %
}}

\def\NewLine{\par\indent}
\def\removebox{\setbox0=\lastbox}

\def\|{|}

\newcommand{\RNAiFold}{\mbox{\tt RNAiFold}\xspace}
\newcommand{\RNAiTFold}{\mbox{\tt RNAiFold2T}\xspace}
\newcommand{\FRNA}{\mbox{\tt Frnakenstein}\xspace}
\newcommand{\SD}{\mbox{\tt SwitchDesign}\xspace}

\newcommand{\NUPACK}{\mbox{\tt NUPACK}\xspace}
\newcommand{\tablefont}[1]{\fontsize{2.0mm}{2.0mm}\selectfont}

\begin{document}

\title{RNAiFold2T: Constraint Programming design of thermo-IRES switches}
\author{Juan Antonio Garcia-Martin\,$^1$,
Ivan Dotu\,$^2$,
Javier Fernandez-Chamorro\,$^3$,\\
Gloria Lozano\,$^3$,
Jorge Ramajo\,$^3$,\\
Encarnacion Martinez-Salas\,$^3$,
Peter Clote\,$^1$\thanks{ Corresponding author: {\tt clote@bc.edu}}
}
\date{\small 
1: Biology Department, Boston College, Chestnut Hill, MA 02467 (USA).\\
2: Research Programme on Biomedical Informatics (GRIB), 
Department of Experimental and Health Sciences, 
Universitat Pompeu Fabra. Dr. Aiguader 88. Barcelona, (Spain). \\
3: Centro de Biologia Molecular Severo Ochoa, 
Consejo Superior de Investigaciones Cientificas -- Universidad 
Autonoma de Madrid, Nicolas Cabrera 1, 28049 Madrid (Spain).
}

\maketitle

\begin{abstract}
\textbf{Motivation:} 
RNA {\em thermometers} (RNATs) are {\em cis}-regulatory elements 
that change secondary structure upon temperature shift. 
Often involved in the regulation of heat shock, cold shock and virulence 
genes, RNATs constitute an interesting potential resource in synthetic 
biology, where engineered RNATs could prove to be useful tools in
biosensors and conditional gene regulation.  \\
\textbf{Results:} 
Solving the 2-temperature inverse folding problem is critical for
RNAT engineering. Here we introduce {\tt RNAiFold2T}, the
first Constraint Programming (CP) and Large Neighborhood Search (LNS) 
algorithms to solve this problem. Benchmarking
tests of {\tt RNAiFold2T} against existent programs 
(adaptive walk and genetic algorithm) 
inverse folding show that our software generates two orders of magnitude
more solutions, thus allowing ample exploration of the space of solutions.
Subsequently, solutions can be prioritized by computing various measures,
including probability of target structure in the ensemble, 
melting temperature, etc. 
%
Using this strategy, we rationally designed two thermosensor
internal ribosome entry site ({\em thermo}-IRES) elements,
whose normalized cap-independent translation efficiency is approximately
50\% greater at 42$^{\circ}$C than 30$^{\circ}$C, when tested in
reticulocyte lysates.  Translation efficiency
is lower than that of the wild-type IRES element,
which on the other hand is fully resistant to temperature shift-up.
This appears to be the first
purely computational design of functional RNA thermoswitches, and 
certainly the first purely computational design of functional 
thermo-IRES elements.  \\
%
\textbf{Availability:}  
{\tt RNAiFold2T} is publicly available as as part of the new
release {\tt RNAiFold3.0} at
\url{https://github.com/clotelab/RNAiFold} and
\url{http://bioinformatics.bc.edu/clotelab/RNAiFold}, which latter
has a web server as well.
The software is written in C++ and uses OR-Tools CP search engine.
\\
\textbf{Contact:} {\tt clote@bc.edu}\\
\textbf{Final version:} This article will appear in {\em Bioinformatics}
2016.
\end{abstract}

\section{Introduction}
RNA {\em thermometers} (RNATs), also known as {\em thermosensors},
are {\em cis}-regulatory elements that change secondary structure upon
temperature shift. Examples include 
(1) repression of heat shock gene expression (ROSE) elements 
\cite{Nocker.nar01},  that control the expression of 
small heat shock genes, such as hspA in {\em Bradyrhizobium japonicum}
and ibpA in {\em Escherichia coli},
(2) FourU elements \cite{Waldminghaus.mm07}, such as the virulence factor
LcrF in {\em Yersinia pestis},
(3) Hsp17 thermosensor \cite{Torok.pnas01,Kortmann.nar11}, which controls 
membrane integrity of the cyanobacterium 
{\em Synechocystis sp. PCC6803} under stress conditions, critical for 
photosynthetic activity. Additional 
examples are described in \cite{Kortmann.nrm12}. 
ROSE elements and FourU elements operate as temperature-sensitive,
reversible zippers, while the {\em Listeria monocytogenes} prfA thermosensor
\cite{Johansson.c02}, phage $\lambda$ cIII thermoswitch \cite{Altuvia.jmb89}
and {\em E. coli} CspA cold shock thermometer 
\cite{Bae.pnas00} operate in a switch-like fashion.
Here, the helix of a zipper melts gradually with increasing temperature,
returning to the original structure when temperature is reduced, while a
switch consists of two mutually exclusive structures determined by
temperature.

Several bioinformatics search methods exist to identify and predict
candidate RNA thermometers. In \cite{Waldminghaus.mgg07} the
database RNA-SURIBA (Structures of Untranslated Regions In BActeria)
was created; using regular expressions, particular structural motifs
were detected in the minimum free energy (MFE) structure, as determined
by {\tt mfold} \cite{zuker:mfoldWebserver}. In contrast, the
{\tt RNAtips} web server 
\cite{Chursov.nar13} and the {\tt RNAthermsw} \cite{Churkin.po14}
web server both rely on base pairing probabilities computed at 
different temperatures using {\tt RNAfold} from the
{\tt Vienna RNA Package} \cite{Lorenz.amb11}.

For some time now, RNA thermosensors have been recognized as
an attractive target for rational design \cite{Wieland.cb07,Waldminghaus.bc08}.
Indeed, within the broader context of synthetic biology, 
rationally designed thermometers could be used as a {\em thermogenetic} 
tool to control expression by temperature regulation 
(i.e. {\em on-demand protein translation}), or even as a multifunctional
nanoscale devices to measure temperature in the context of 
hyperthermic treatment of cancer cells, imaging, or drug delivery
\cite{nanoscaleThermometer}.

In \cite{Neupert.nar08}, synthetic (zipper) thermosensors 
were {\em manually} designed
to sequester the Shine-Dalgarno (SD) sequence AAGGAG within a single
stem-loop structure containing 
4-9 base pairs, several of which contained 1-2  bulges of size 1.
In \cite{Waldminghaus.bc08}, synthetic (switch) thermosensors were 
{\em computationally} designed to switch between a single  
stem-loop structure that sequesters the SD sequence GGAGG, 
and two shorter stem-loop structures where the SD sequence is found in
the apical loop of the second stem-loop. In particular, the 2-temperature
inverse folding (adaptive walk) program \SD  \cite{flamm:RNAswitch} 
was used to obtain 300 candidate
sequences; only two candidate sequences survived after the application of
several computational filters including the computation of melting
curves with {\tt RNAheat} \cite{Lorenz.amb11}.  Since neither 
of these sequence displayed
any temperature-dependent control of a reporter gene (bgaB) fusion,
the top candidate sequence was used as a template in
two rounds of error-prone mutagenesis followed by selection, resulting in
a successful thermosensor  -- see Figure~5 of \cite{Waldminghaus.bc08}.
In \cite{Ogawa.r11}, a non temperature-dependent {\em riboswitch} was
manually designed, which promotes cap-independent translation in 
wheat germ cell lysate only upon binding of the ligand theophylline.
In \cite{Wachsmuth.nar13}, a synthetic theophylline riboswitch was designed
by a computational pipeline including inverse folding
and experimentally shown to perform transcriptional regulation in
{\em Escherichia coli} Top10 cells.
In \cite{Saragliadis.rb13}, a {\em thermozyme} was created by fusing a
{\em Salmonella} RNA thermometer (RNAT) to a hammerhead
ribozyme, followed by {\em in vivo} screening -- thus showing that
naturally occurring hammerheads and RNATs appear to be modules that can
be combined.
In \cite{HoynesOConnor.nar15} small, heat-repressible  RNA 
thermosensors (zippers) were manually designed in {\em E. coli}, which at 
low temperature sequester a cleavage site for RNaseE, and at high temperatures
unfold to allow mRNA degradation.

Despite these impressive results, \cite{Kortmann.nrm12} state that:
``RNATs have little, if any sequence conservation and are difficult to 
predict from genome sequences. \ldots
Therefore, the bioinformatic prediction and rational design of 
functional RNATs has remained a major challenge''.


In this paper, we introduce the software {\tt RNAiFold2T}, capable of 
solving the inverse folding problem for two or more temperatures, i.e.
generating one or more RNA sequences whose minimum free energy (MFE)
secondary structures at temperatures $T_1$ and $T_2$ [resp. $T_1,\ldots,T_m$]
are user-specified target structures $S_1$ and $S_2$ [resp. $S_1,\ldots,S_m$], 
or which reports that no such solution exists. {\tt RNAiFold2T} is unique
in that it implements two different algorithms -- {\em Constraint Programming} 
(CP) and {\em Large Neighborhood Search} (LNS). CP is an exact, non-heuristic
method that uses an exhaustive yet efficient branch-and-prune process, and
is the only currently available software capable of generating all solutions
or determining that no solution exists (since there are possibly exponentially
many solutions, a complete solution is feasible only for structures of modest
size). LNS uses a local search heuristic, complemented with local calls of
constraint programming to explore solutions of substructures of the target
structures.
We use {\tt RNAiFold2T} to rationally design two {\em thermo}-IRES elements,
whose normalized cap-independent translation efficiency is approximately
50\% greater at 42$^{\circ}$C than 30$^{\circ}$C, when tested in
reticulocyte lysates.  
We then benchmark {\tt RNAiFold2T} with the only two other programs
that solve the 2-temperature inverse folding problem: the adaptive walk
\SD ({\tt SD}) \cite{flamm:RNAswitch}  and  the genetic algorithm
\FRNA ({\tt FRNA}) \cite{frnakenstein}. {\tt RNAiFold2T} CP
generates two orders of magnitude more solutions than either, when all
programs are run for 24 hours, while the number of distinct 
problem instances that can be solved by {\tt RNAiFold2T} LNS
within 30 [resp. 60] minutes for short [resp. long] target structures 
is roughly comparable for each program.  Finally, by analyzing
existent RNATs found in the Rfam 12.0 database \cite{Nawrocki.nar14},
we determine that naturally occurring RNATs appear not to be optimized for
the {\em cost} function (equation (7) of \cite{flamm:RNAswitch}), and that both
{\tt SD} and {\tt FRNA} appear to generate solutions 
whose {\em cost} function value
is substantially lower than that of naturally occurring RNATs, in contrast
to solutions returned by {\tt RNAiFold2T}.

\section{Methods}
\subsection{Biochemical methods}

{\bf\em Thermo-IRES activity assay:}
Thermo-IRES constructs were created by replacement of wild-type nucleotides
at positions 417-462 in domain 5 of the foot-and-mouth disease virus
(FMDV) IRES element, whose 
secondary structure is depicted in Figure~1 of \cite{Fernandez11a}
and Figure~2 of \cite{Lozano.fj14}.
Six computationally designed thermo-IRES elements were selected for validation,
along with a negative control and the wild-type FMDV IRES element as 
positive control. 
Specifically, synthetic oligonucleotides containing the designed sequence 
(46 nts) in either positive or negative orientation were annealed in
Tris 50 mM pH 7.5, NaCl 100 mM, MgCl$_2$ 10 mM, 15 min at 37$^{\circ}$C
and subsequently inserted into the HindIII and XhoI restriction sites of 
pBIC, which harbors the wild-type IRES, linearized with the same enzymes. 
Colonies that carried the correct insert were then selected, and prior
to expression analysis, the nucleotide sequence of the entire length of 
each region under study was determined (Macrogen).

{\em In vitro} transcription was performed for 1 h at 37$^{\circ}$C
using T7 RNA polymerase, as described in \cite{Lozano.rb15}.
RNA was extracted with phenol-chloroform, ethanol precipitated and then
resuspended in TE. 
Using gel electrophoresis, the transcripts were checked for integrity.
Equal amounts of the RNAs synthesized in vitro were translated in 
70\% rabbit reticulocyte lysate (RRL) (Promega) supplemented with 
$^{35}$S-methionine (10 $\mu$Ci), as described in \cite{Pineiro.nar13}.
Each experiment was independently repeated in triplicate,
using the wild type RNA as a control in all assays. 
Luciferase (LUC) and chloramphenicol acetyl transferase (CAT)
activities were measured for the bicistronic plasmid,
as previously described \cite{FernandezChamorro.nar14}. In particular,
intensity of the LUC band, as well as the CAT band,
produced by each transcript 
was determined in a densitometer, and normalized against the intensity of 
LUC and CAT bands produced by the wild type RNA, set at 100\%. 
Values represent the mean $\pm$ SD.

Luciferase activity reflects the efficiency of IRES-dependent translation, 
while CAT activity reflects the efficiency of $5'$-dependent translation; 
thus the ratio LUC/CAT
was determined at 30$^{\circ}$C and 42$^{\circ}$C  for
wild-type FMDV IRES, the negative control and two thermo-IRES constructs.
Selective $2'$-hydroxyl acylation analyzed by primer extension (SHAPE)
was performed for wild-type IRES element, as described in
\cite{Dotu.rb13}
\smallskip

\subsection{Computational methods}

{\tt RNAiFold2T} uses Constraint Programming (CP) to determine those 
sequences, whose minimum free energy (MFE) structure at temperature $T_1$ 
[resp. $T_2$] is identical to a user-specified target structure $S_1$
[resp. $S_2$].  The target structures $S_1,S_2$ can also be hybridization 
complexes of two RNAs, rather than single secondary structures.
CP performs a complete, exhaustive (branch and prune) exploration of 
the search space and therefore, it can return all possible solutions of the 
thermoswitch design problem or prove that no solution exists (given 
an unlimited amount of time).
In addition to CP, {\tt RNAiFold2T} also supports Large Neighborhood
Search (LNS),
a fast local (not complete) search metaheuristic that employs CP to 
exhaustively explore
large neighborhoods of every candidate solution at each iteration step.
Moreover, since it is written in C++ using
the OR-Tools engine \url{https://github.com/google/or-tools}, together
with plug-ins to {\tt Vienna RNA Package} \cite{Lorenz.amb11} and
{\tt RNAstructure} \cite{Reuter.bb10}, the user 
can install and run {\tt RNAiFold2T} locally, thus permitting much longer
execution times than supported by our web server.

The overall methodology of {\tt RNAiFold2T}
is similar to its precursor, {\tt RNAiFold 2.0}
\cite{GarciaMartin12,GarciaMartin.nar15}; however, as explained below,
there are a number of algorithmic details that are new and not present
in {\tt RNAiFold 2.0} -- decomposition tree for 2 or more target structures,
novel constraints that are underlined below,
variable (helix) and value heuristics that are proper only to {\tt RNAiFold2T},
the introduction of two types of restart heuristic in LNS
to ensure a good trade-off between exploration of promising regions of 
the search space versus the exploration of remote portions of the search space. 
{\tt RNAiFold2T} cleanly separates all constraints 
from the CP or LNS solver, thus permitting our software
to be extended to support any future desirable constraints. 
Current constraints supported by {\tt RNAiFold2T} 
include the following (novel constraints underlined): \underline{{\em full} 
and/or {\em partial}
target structures or hybridization complexes at two temperatures};
\underline{a plug-in to determine MFE structure using either {\tt RNAfold}}
\cite{Lorenz.amb11} \underline{or {\tt RNAstructure}} \cite{Reuter.bb10} 
\underline{with
the Turner99, Turner2004} \cite{Turner.nar10} \underline{or Andronescu}
\cite{Andronescu.b07} \underline{energy model} with dangle treatment (-d0, -d1, -d2, -d3
corresponding respectively  to no dangle, max of $5'$ and $3'$ dangle, 
sum of $5'$ and $3'$ dangle, sum of $5'$ and $3'$ dangle with coaxial stacking);
IUPAC nucleotide constraints, IUPAC amino acid constraints
that require all returned RNA sequences to code specified peptides in
one or more overlapping reading frames, structural compatibility and
structural incompatibility constraints, etc. Additionally, {\tt RNAiFold2T}
can determine a user-specified number of solutions, all
solutions (given sufficient time), or whether no solution exists. Indeed,
memory requirements for {\tt RNAiFold2T} are minimal, and since there are
no memory leaks,  the software can be run for weeks.

In developing a CP solution to a given problem the main tasks are 
to define the problem (specify variables, domains and constraints) and 
to define the search procedure. The extension of the model from RNAiFold2.0 
is trivial and consists of adding new constraints for the helices 
corresponding to the second 
structure. The search procedure, however, must be adapted to the 
new difficulties imposed by the 2-temperature problem. New variable 
and value ordering heuristics are needed in order to solve 
2-temperature inverse folding efficiently 
(see Supplementary Tables 1,2). The algorithmic details 
related to the new search procedure are explained below:
		
\subsubsection{Structure decomposition}	

As in other inverse folding methods, such as
{\tt RNAinverse} \cite{Lorenz.amb11},
{\tt NUPACK} \cite{Zadeh.jcc11},
etc.,  we rely on a 
decomposition tree of the structure into independent helices, called
{\em extended helices} (EHs) and {\em extended helices with dangles}
(EHwDs). See Supplementary Information (SI) for definitions of EH, EHwD and
decomposition tree, and see Figure~1 for an illustration of the EHwD
decomposition tree for a FourU thermometer.  Decomposition trees play
a special role in {\tt RNAiFold2T} for the following reasons:
(a) each node in the decomposition tree is a constraint,
(b) the helix and variable heuristics (described later) cause 
the search tree to be searched in a specific order. 
To improve efficiency in solving multi-temperature inverse folding, 
we investigated various helix and value heuristics, which 
steer the search within a search space defined by a composite decomposition
tree, comprising subtrees for each target structure at the corresponding 
temperature.  The helix and value heuristics are new and not part of 
{\tt RNAiFold 2.0}.

\begin{figure}[htbp]
\centering
\begin{subfigure}[b]{0.44\textwidth}
\includegraphics[width=\textwidth]{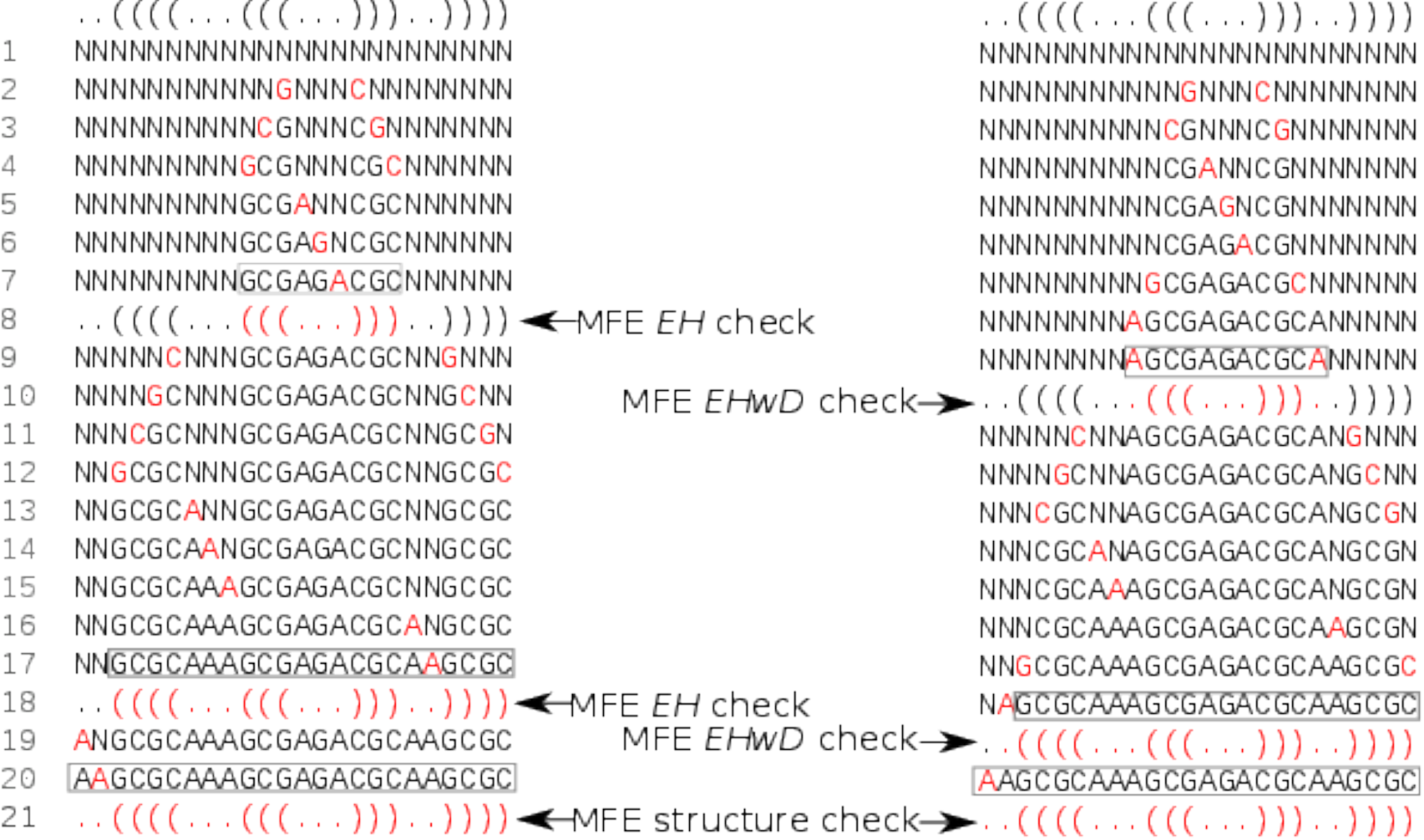}
\caption{\small EH (left) and EHwD (right)} 
\label{fig:1a}
\end{subfigure}%
\qquad
\begin{subfigure}[b]{0.44\textwidth}
\includegraphics[width=\textwidth]{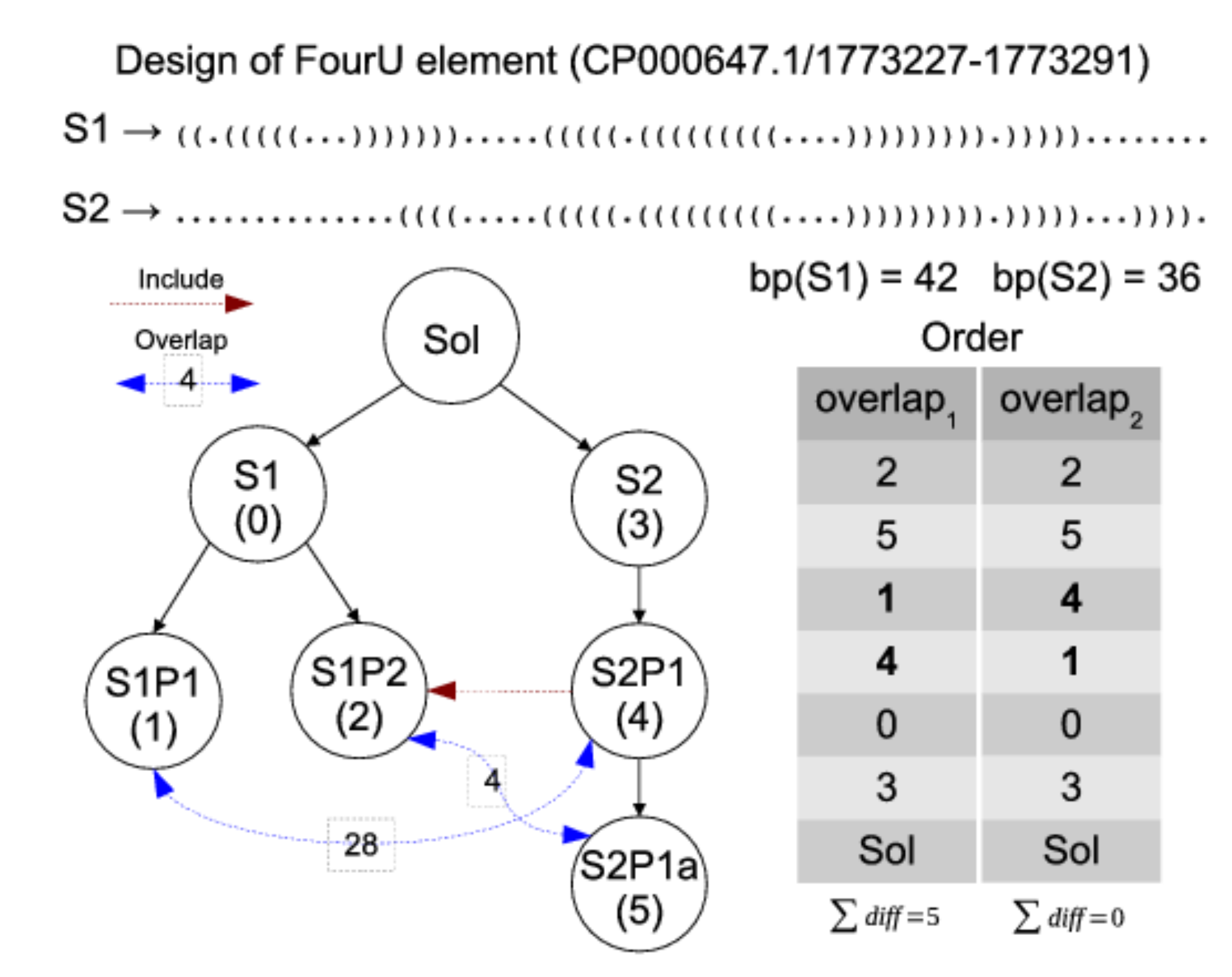}
\caption{\small EHwD tree decomposition}
\label{fig:1b}
\end{subfigure}
\caption{\small Examples of CP search for helixes vs EHwDs and EHwDs decomposition heuristic. 
{\em (a)} Minimum free energy (MFE) structure check made for extended helices
(EH) on the left, and extended helices with dangles ((EHwD) on the right.
In each line, either a new base pair (2 positions) or a new nucleotide
(1 position) is instantiated, following the {\em variable order heuristic}.
For simplification, this example assumes that correct value  assignments 
are made in each instantiation step.
{\em (b)} Tree decomposition for target structure $S_1$ [resp. $S_2$]
at temperature $T_1$ [resp. $T_2$], using target structures for the
FourU element CP000647.1/1773227-1773291. The table on the right presents
the order of helix exploration using $overlap_1$ and $overlap_2$ heuristics.
}
\label{fig:1}
\end{figure}

Consider the 65 nt FourU RNA thermometer
whose MFE structures determined by
{\tt RNAfold} from {\tt Vienna RNA Package 2.1.9} \cite{Lorenz.amb11} at
37$^{\circ}$C and 53$^{\circ}$C are given in dot-bracket notation  by
\begin{tiny}
\begin{verbatim}
>CP000647.1/1773227-1773291)
12345678901234567890123456789012345678901234567890123456789012345
GGACAAGCAAUGCUUGCCUUUAUGUUGAGCUUUUGAAUGAAUAUUCAGGAGGUUAAUUAUGGCAC
((.(((((...))))))).....(((((.(((((((((....))))))))).)))))........
..............((((.....(((((.(((((((((....))))))))).)))))...)))).
\end{verbatim}
\end{tiny}
Let $S_1$ [resp. $S_2$] denote the MFE structure of this FourU thermometer
at 37$^{\circ}$C [resp. 53$^{\circ}$C].
$S_1$ is identical to the consensus structure 
from Rfam 12.0 \cite{Nawrocki.nar14}, 
as well as the structure displayed in Figure~1
of \cite{Waldminghaus.mm07} for the FourU sequence taken from the 
$5'$-UTR of the {\em Salmonella} agsA gene. 
Giving labels as described to nodes in the 
EHwD decomposition trees $\mathcal{T}_1,\mathcal{T}_2$  respectively
for $S_1$ and $S_2$, we find
the EHwD decomposition of $S_1$ has EHwD 0 from positions 1-65,
EHwD 1 from positions 1-19, and EHwD 2 from positions 23-58, while the
EHwD decomposition of $S_2$ has EHwD 4 from positions 1-65, 
EHwD 5 from positions 14-65, and EHwD 5 from positions 23-58.
The lower portion of 
Figure~\ref{fig:1} depicts the EHwD decomposition trees for
temperatures $T_1,T_2$, joined together with a (dummy) root 
that corresponds to the solution returned by {\tt RNAiFold2T}.

It can happen that the MFE structure of an extended helix
does not agree with the target substructure, while that of the extended helix
with dangles does. This is the reason that the
EH decomposition trees of {\tt RNAiFold 1.0} \cite{GarciaMartin12}
were replaced by 
EHwD decomposition trees in {\tt RNAiFold 2.0} \cite{GarciaMartin.nar15}, 
and why EHwD trees are used in {\tt RNAiFold2T}.

In the CP and LNS search strategy, 
whenever a subsequence corresponding to a node
of the decomposition tree has been instantiated, a check is made to determine
whether the MFE structure of the subsequence is identical to the target
substructure. In the case of structural disagreement, the instantiated 
subsequence is discarded and backtracking occurs. For any solution sequence
returned by {\tt RNAiFold2T}, it follows that at temperature $T_1$ 
[resp. $T_2$], each subsequence of the solution that corresponds to a node in
decomposition tree  ${\mathcal{T}}_1$ [resp. ${\mathcal{T}}_2$]  folds into the
corresponding substructure of target $S_1$ [resp. $S_2$].
The top portion of Figure~\ref{fig:1} illustrates the differences
between extended helix (EH) and extended helix with dangle (EHwD) 
decomposition, and when a check is made for whether the target substructure
agrees with the  MFE structure of the instantiated subsequence.
In the sequel, we refer to four types of elements in an EHwD:
\begin{itemize}
\item Dangling position: Unpaired position at any side of a helix.
\item Unpaired position: Any other unpaired position.
\item Closing base pair: Outermost base pair of a helix.
\item Normal base pair: Any other base pair.
\end{itemize}

\subsubsection{Heuristics for variable and value order}

In a Constraint Programming (CP) algorithm, one typically specifies 
the order in which variables are instantiated (assigned), known as
the {\em variable ordering heuristic}, as well as the order in which the
values belonging to the domain of each variable are to be assigned, known
as the {\em value ordering heuristic}. 
The variable ordering heuristic is divided into
two levels: first, the order in which extended helices with dangles
(EHwDs) are to be assigned, and
second, the order in which nucleotide positions within helices are to be
assigned.  In this
section, we also discuss {\em restarting heuristics} for the 
Large Neighbourhood Search (LNS) variant of {\tt RNAiFold2T}.

\paragraph*{Variable ordering at the helix level:}
In the search for thermosensors, there is often an overlap between EHwDs
of structure $S_1$ and those of structure $S_2$ -- this situation 
substantially complicates
the task of finding an optimized order of exploration of the CP search space.
In the leaf-to-root heuristic of {\tt RNAiFold 2.0}, 
EHwD node $H$ is explored before EHwD node $H'$ if the
height $ht(H)$ of $H$ is less than the height $ht(H')$ of $H'$,
or if $ht(H)=ht(H')$ and $H$ appears to the left of $H'$ in the
decomposition tree for the single target structure $S$.
In contrast to this heuristic,
\RNAiTFold implements two different approaches in order to find an adequate 
exploration ordering for the extended helices with dangles for two target
structures $S_1$ and $S_2$, whereby a high priority is given to solve 
those helices, whose sequence is determinant for other parts of the 
structure due to overlaps.  Let $N$ denote the number of
nodes (EHwDs) in the decomposition tree for $S_1$, plus the number  of nodes
(EHwDs) in the decomposition tree for $S_2$. Suppose that $H,H'$ are two
distinct EHwDs belonging to $S_1$ or $S_2$, where the outermost base pair
of $H$ [resp. $H'$] is $(i,j)$ [resp. $(i',j')$]. Define the following
relations:

\begin{itemize}	
\item 
$includes(H,H')$ is $1$ if
$i<i'$ and $j>j'$, i.e. interval $[i,j]$ properly contains 
$[i',j']$; otherwise $includes(H,H')$ is $0$.

\item
$overlap_1(H,H')$ is $1$ if $[i,j] \cap [i',j'] \ne \emptyset$,
or equivalently $\max(i,i') \leq \min(j,j')$; otherwise 
$overlap_1(H,H')$ is $0$.

\item 
$overlap_2(H,H')$ is the number of positions $k$ in
$H$ and $H'$, for which $k$ is base-paired in both $H$ and $H'$.

\item 
$degree_{\alpha}(H) = \sum_{i=0}^N overlap_{\alpha}(H, H_i)$,
for $\alpha=1,2$.

\item 
For $\alpha=1,2$,
$degreedist_{\alpha}(H,H')$ is equal to
$degree_{\alpha}(H)- degree_{\alpha}(H')$, provided that
$degree_{\alpha}(H) > degree_{\alpha}(H')$;
otherwise, $degreedist_{\alpha}(H,H')$ is  $0$.
	
\item 
For $\alpha=1,2$,
$diff_{\alpha}(\sigma,H,H')$ is equal to
$degreedist_{\alpha}(H,H')$, provided that
$\sigma(label(H)) < \sigma(label(H'))$; 
otherwise
$diff_{\alpha}(\sigma,H,H')$ is $0$.

\end{itemize}
The value $label(H)$ is defined in SI, and corresponds to
visitation order in breadth-first traversal of tree $\mathcal{T}$, and
$\sigma: \{ 0,\ldots,N-1 \} \rightarrow \{ 0,\ldots,N-1 \}$ is
a permutation that minimizes
$\sum_{i=0}^N\sum_{j=0}^N diff_{\alpha}(\sigma,H_i, H_j)$,
subject to the constraint that if $H$ properly includes $H'$,
then $order(H) > order(H')$. Note that the partition $\sigma$ orders
the EHwDs of $S_1$ and $S_2$ in order to minimize the total
{\em incremental overlap}.
Before the search for thermosensors begins, {\tt RNAiFold2T} executes a very
fast CP search to determine the optimal ordering permutation $\sigma$.
Finally, by setting the index $\alpha$
to either $1$ or $2$ in the definition of 
$diff_{\alpha}(\sigma,H,H')$,
we obtain the first or second search heuristic. (Two additional
helix ordering heuristics are explained in Supplementary Information.)
An example of each helix order heuristic $1,2$ is shown in 
Figure~\ref{fig:1} -- note that for even small structures,
there are several differences in the helix exploration order when
$\alpha$ is $1$ or $2$.


\paragraph*{Variable ordering at the nucleotide level:}
The second level of variable ordering heuristic deals with the 
exploration of nucleotide positions within a given EHwD structure. 
This second level of variable ordering can be 
stated as follows:

\begin{itemize}
\item First, non-outermost base-paired positions $(x,y)$ of a given EHwD 
are instantiated from the innermost base pair to the outermost base pair.

\item Second, unpaired positions in a given EHwD are grouped together
in consecutive runs, and these runs are ordered from largest to smallest
and then instantiated from left to right.

\item Third, the outermost, closing base pair of a given EHwD is
instantiated.

\item Finally, dangling positions of a given EHwD (if any) are instantiated 
(note that not all EHwDs contain dangling positions).
\end{itemize}

\begin{figure}[htbp]
\includegraphics[width=0.9\textwidth]{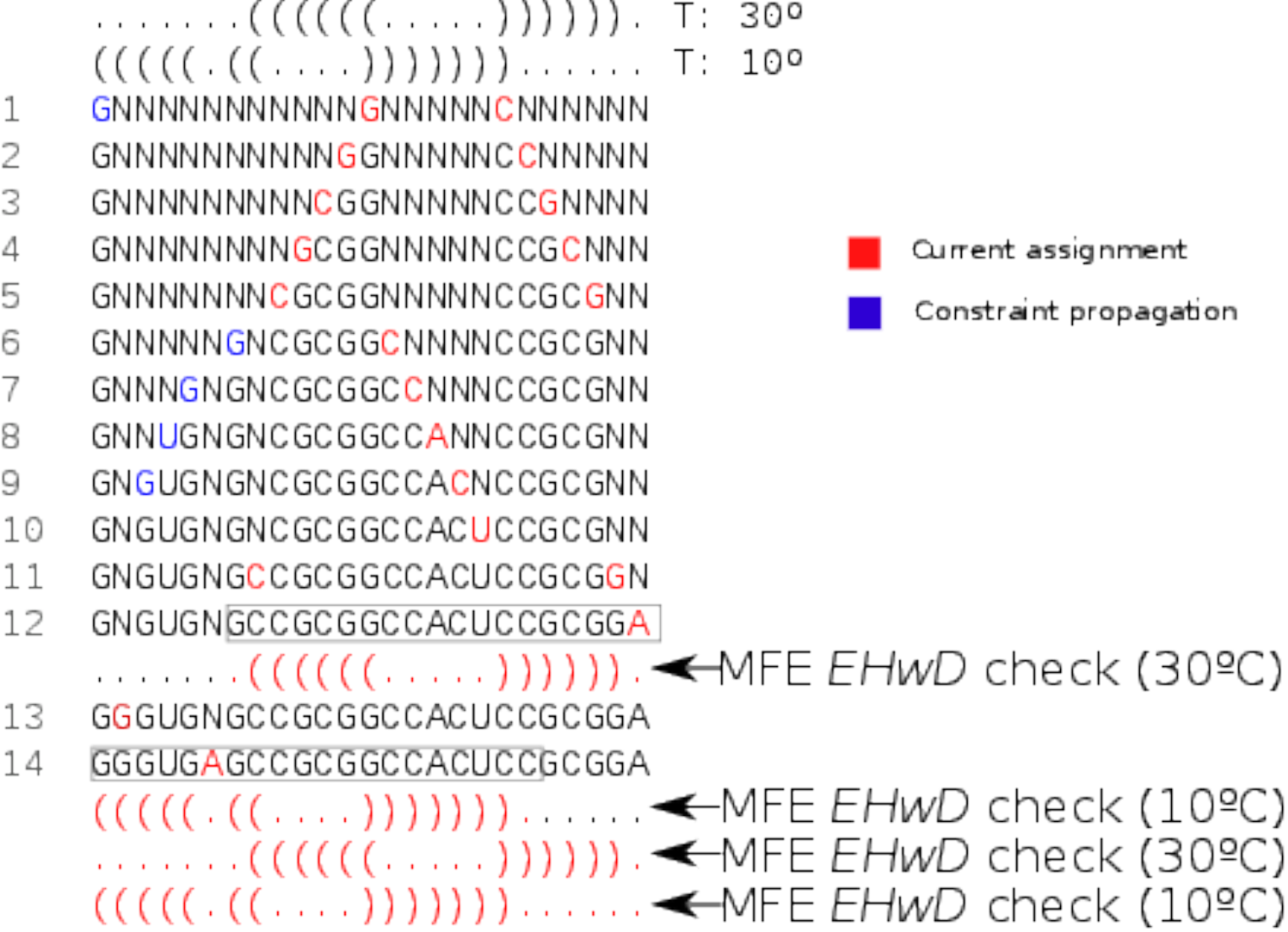}
\caption{\small Example of CP search for target structures $S_1$ (top) at temperature
30$^{\circ}$C, and $S_2$ (bottom) at temperature 10$^{\circ}$C.
Undetermined positions are assigned (red) 
in the following order: base pairs are instantiated in lines
1,2,3,4,5,13; unpaired positions in lines
6,7,8,9,10,14;  a closing base pair in line 11;
a dangling position in line 12.
When the left nucleotide of a base pair is instantiated to C [resp. A],
then propagation of the base pairing constraint reduces the domain of
possible values of the right nucleotide of the base pair from 
$\{ A,C,G,U \}$ to the set $\{ G \}$
[resp. $\{ U \}$] -- this happens in lines 1,6-9. 
For simplification, this example assumes that
correct value  assignments always occur at each search step.
}
\label{fig:2}
\end{figure}

The small example shown in Figure~\ref{fig:2} 
graphically depicts the differences between constraint checks and 
variable ordering for both \RNAiFold and \RNAiTFold. 
Supplementary Tables 1,2 describe and benchmark variable and value
heuristics in {\tt RNAiFold2T}.

\paragraph*{Value ordering:}
Value ordering establishes the order in which values are
assigned to variables -- in our case, this means values
{GC,CG, AU, UA, GU, UG} for base-paired positions and
{A,C,G,U} for unpaired positions. The underlying idea for ordering
domain values for variables for base-paired positions and unpaired positions
is to allow the creation of thermodynamically 
stable helices and to take into account the nature of the overlap in
overlapping positions. This requires specific value orderings for
base-paired positions, depending on whether the position is a dangle,
mismatch, normal or closing base pair of an EHwD for both targets $S_1,S_2$,
or only one of the target structures.  For each of these cases, we define
different ordering heuristics, described in Supplementary Table 1.

\paragraph*{LNS restart heuristics:}	
Similarly as in the LNS variant of \ RNAiFold, the restart condition 
is a given amount of time, proportional to the length of the target 
structures, after which search is stopped and some variables are fixed 
in order to explore exhaustively a large neighbourhood of the current 
solution.  After the first restart, full exploration of the remaining 
space with no solution found is also a restart condition.

In \RNAiTFold, when a restart is triggered, a set of positions 
is selected as candidates to be fixed. The MFE structure for each 
EHwD of the current candidate solutions is evaluated independently. 
If the MFE structure of an EHwD matches with the target structure 
at the desired temperature, and the MFE structure of 
all the overlapping helices in the second target structure (at the 
corresponding temperature) also matches, then all the EHwD positions 
are included in the set of candidates. 
When all the EHwDs have been evaluated, candidate positions are fixed 
with a probability of $0.9$, and the set of candidate positions is stored.

Since the order of exploration is similar in each round, it could be 
possible that fixing similar parts of the sequence results in an 
exploration of almost the same region of the search space in subsequent 
searches, so two mechanisms are implemented to avoid this behavior: 
(1) In subsequent restarts, if the candidate positions to be fixed are the 
same as in the previous restart, then the probability of fixing positions 
decreases by $0.05$, if not, then the initial probability of $0.9$ is restored.
(2) There is a hard restart (no nucleotide position is fixed) in the case
that, after $10$ restarts, the set of candidate positions remains unchanged, 
or if all possible solutions for the current subproblem have been explored.

In local search algorithms, there is always a trade-off between 
{\em exploitation} and {\em exploration}.
Exploitation means focusing the search on promising regions, as reflected in
our choice of probability $0.9$ to remain close to currently instantiated
portions of the 
sequence. Exploration means covering different, remote regions of the search 
space, as reflected in our choice to decrement the probability by $0.05$ 
and our choice to perform a hard restart after 10 restarts.

\subsection{Benchmarking}

\paragraph*{Benchmarking data:}
We retrieved the sequences of seven families of non-coding RNA 
thermometers from Rfam:
RF00038, RF00433, RF00435, RF01766, RF01795, RF01804, RF01832.
These families include both cold and heat shock RNA thermometers, taken
from diverse organisms including phages, prokaryotes and eukaryotes, with
sequence length ranging from 60 nt to 450 nt. The benchmarking 
was divided into two groups: sequences shorter than and longer than 130 nt.
For each sequence, we used {\tt RNAfold} \cite{Lorenz.amb11}
with the Turner99 energy parameters \cite{Turner.nar10} to determine the
MFE structures at temperatures $T_1$ and $T_2$, where temperatures were
chosen (essentially) according to published experimental studies for each 
thermometer family \cite{Rinnenthal.nar11,ROSEswitch,Altuvia.jmb89}
-- in particular, we increased the temperature difference $T_2-T_1$ from the 
published values to ensure that {\tt RNAfold} produced distinct structures
at $T_1$ and $T_2$ if possible (see SI). Turner99 rather than Turner2004 energies were
used, since it required less distortion from published temperatures 
$T_1,T_2$.  All sequences, whose MFE structures at $T_1$
and $T_2$ were identical, were subsequently removed. 
See Supplementary Tables 3-5 and SI Excel files for benchmarking results,
and Supplementary Table 6 [resp. SI Excel file] for a list of
sequences, structures, and temperatures for sequences of length less than
130 nt [resp. greater than 130 nt].
The resulting benchmarking set includes all 5 
{\em Lambda} phage CIII thermoregulator elements (Lambda\_thermo), all
3 FourU thermometers, 11 of 13  repression of heat shock (ROSE) elements,
8 of 14 sequences from a second family of repression of heat shock (ROSE\_2)
elements, 3 of 13 thermoregulators of PrfA virulence genes (PrfA), 
4 of 6 HSP90 {\em cis} regulatory elements (HSP\_CRE), and 14 of 15 
cold shock protein regulator sequences (CspA).

\paragraph*{Software used in benchmarking tests:}
\SD ({\tt SD}) \cite{flamm:RNAswitch}, 
\FRNA ({\tt FRNA}) \cite{frnakenstein}, 
and \RNAiTFold were benchmarked on Rfam family 
RF01804 of Lambda phage CIII thermoregulator elements to determine the
maximum number of distinct solutions over 24 hours,
restarting when no new solution is found within 1 hour. 
Since neither {\tt FRNA} nor {\tt SD} output more than one solution, we
made the following modifications of each program. The genetic algorithm
\FRNA was run as many times as possible over 24 hours 
(each time with a time limit of 1 hour); we then output all sequences 
found in the most recent (internally stored) population which fold into
the target structures $S_1,S_2$ at temperatures $T_1,T_2$.
{\tt SD} returns a single sequence which minimizes a cost function
described in \cite{flamm:RNAswitch}; thus
we modified {\tt SD} source code in order to test 
whether any sequence explored in the search was a solution. 
Sequences were checked at two different points in {\tt SD}: 
when a new sequence is generated by a single mutation ({\tt SD} update),
and when a sequence is selected by minimization of the cost function 
({\tt SD} selected). 
In all cases, \SD was restarted if no new sequence was found in one hour.

For each solution set obtained, additional solutions were generated by 
testing all single point mutations of any solution returned. 
Additionally, a reference solution set was produced by 
running {\tt RNAiFold2T} for several days. 


\section{Results}
\subsection{Design of thermo-IRES switches}

\begin{figure}[htbp]
\centering
\begin{subfigure}[b]{0.16\textwidth}
\includegraphics[width=\textwidth]{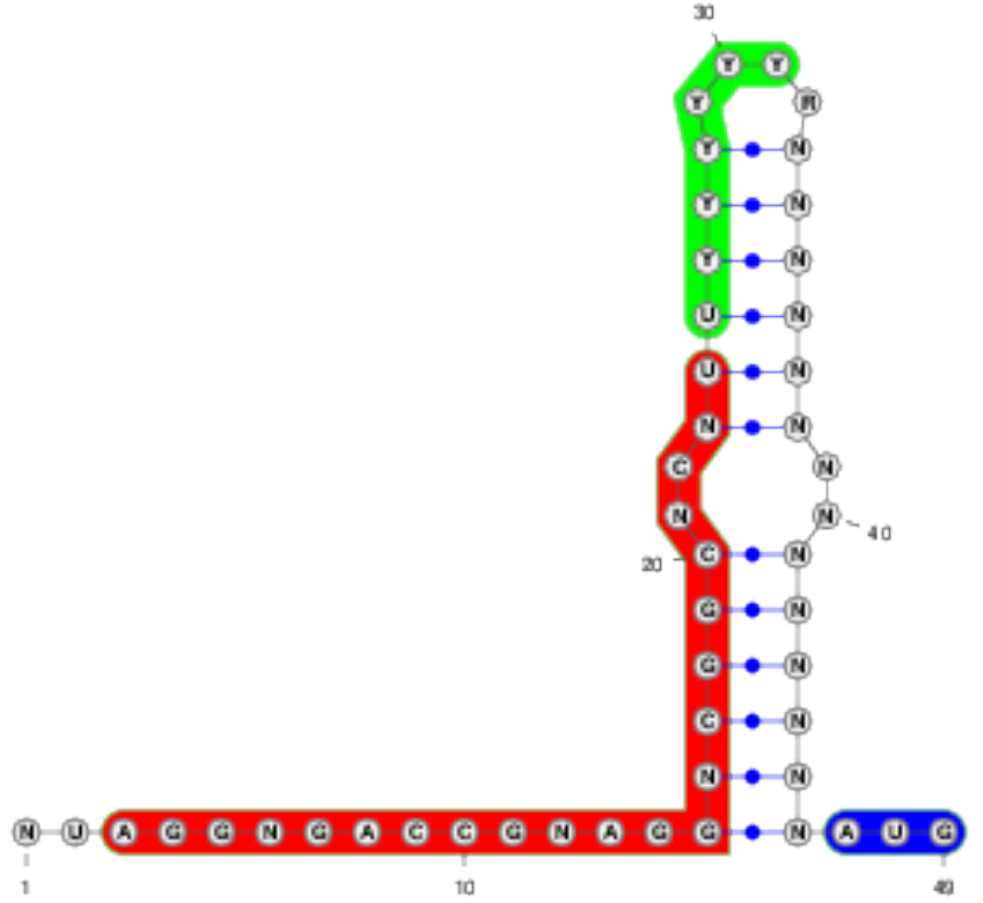}
\caption{}
\label{fig:thermoIREStarget1}
\end{subfigure}%
\quad
\begin{subfigure}[b]{0.25\textwidth}
\includegraphics[width=\textwidth]{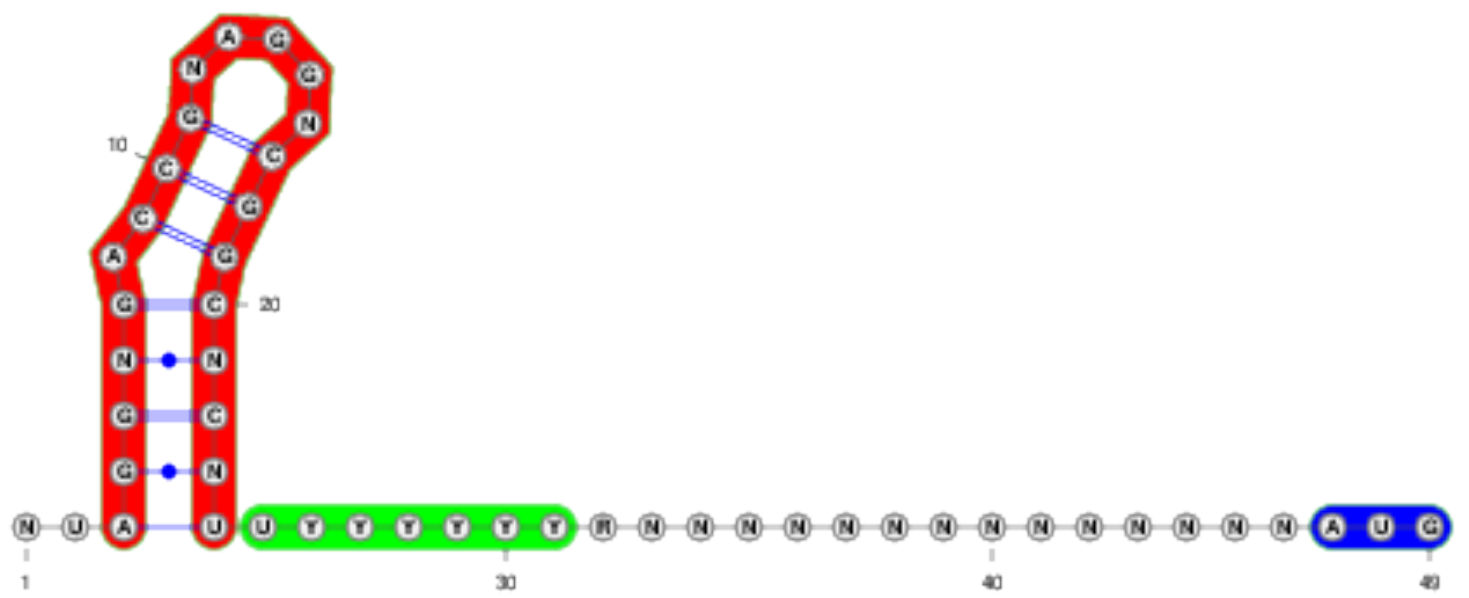}
\caption{}
\label{fig:thermoIREStarget2}
\end{subfigure}
\\
\begin{subfigure}[b]{0.25\textwidth}
\includegraphics[width=\textwidth]{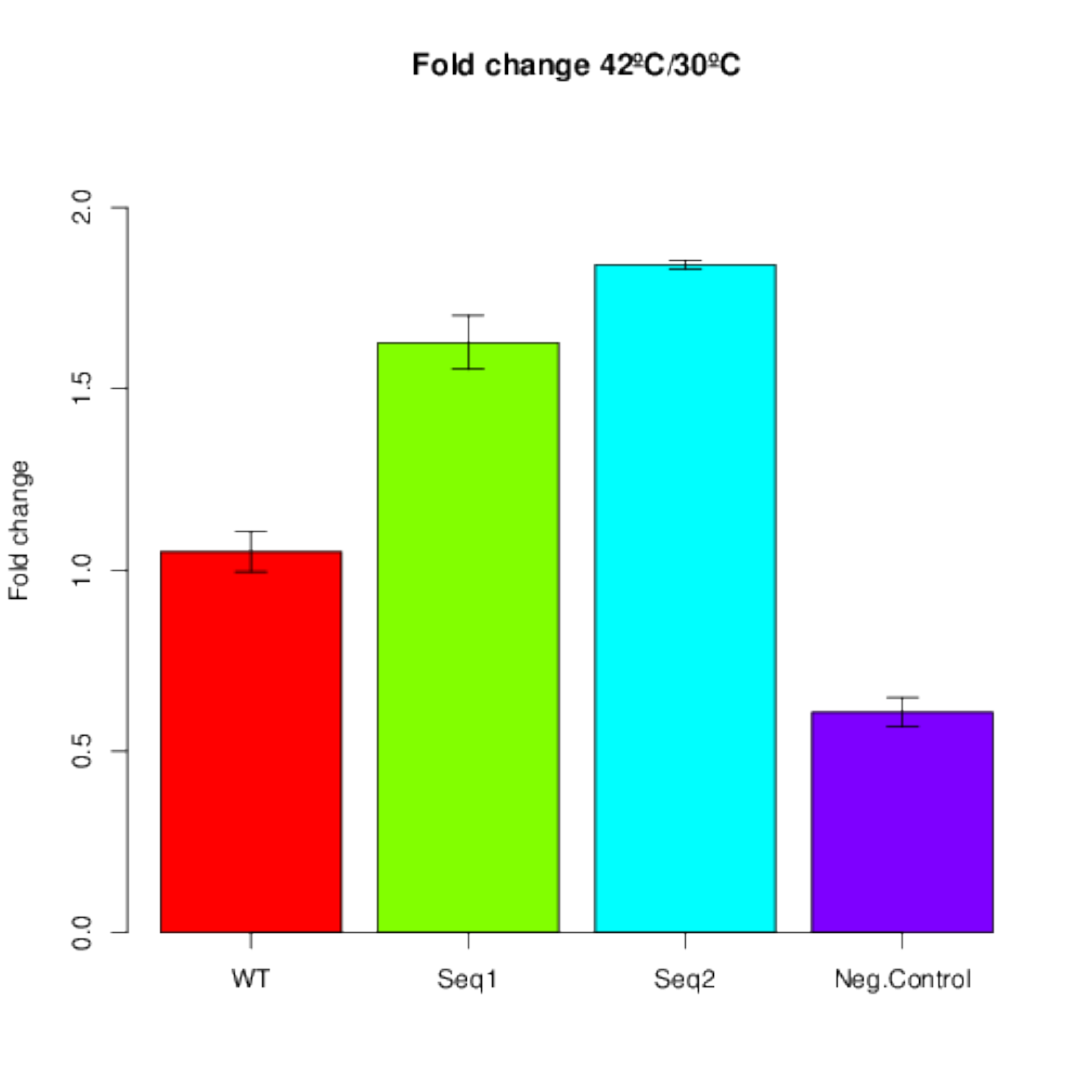}
\caption{}
\label{fig:experimentalResults}
\end{subfigure}
\caption{\small 
{\em (a,b)}
Target structures $S_1$ at temperature 30$^{\circ}$C (a) and
$S_2$ at temperature 42$^{\circ}$C (b) for domain 5 thermo-IRES 
element with added AUG codon (blue). IUPAC sequence constraints are
determined from an alignment of 183 IRES sequences as shown in 
Figure~1 of \cite{Fernandez11a}.
Domain 5 stem-loop (positions 3-24 in red) 
and {\em unpaired} polypyrimidine tract (PPT positions 25-32 in green) are known
to be {\em essential} for IRES activity \cite{Kuhn.jv90}. Target structure
$S_1$ was designed to sequester the  PPT at low temperature, thus creating
a thermo-IRES which should be functional only at high temperature.
{\em (c)}
Ratio of normalized IRES activity at 42$^{\circ}$C over that
at 30$^{\circ}$C for wild-type FMDV IRES, a negative control,
and two thermosensors designed
using {\tt RNAiFold2T}. 
}
\label{fig:thermoIREStarget}
\end{figure}

Domain 5 of wild-type  FMDV IRES element contains the 46 nucleotides
AUAGGUGACC GGAGGUCGGC ACCUUUCCUU UACAAUUAAU GACCCU
at positions 417-462, as shown in Figure~1 of \cite{Fernandez11a} and
Figure~2 of  \cite{Lozano.fj14}. The
domain 5 stem-loop at positions 419-440 and {\em unpaired}
polypyrimidine tract (PPT) region at positions 441-447,
are both known to be essential for IRES activity \cite{Kuhn.jv90}. 
Using the following pipeline, two candidate thermo-IRES elements 
were tested, along with a negative control and a positive control
(wild-type IRES).
\begin{enumerate}
\item
As shown in Figures~\ref{fig:thermoIREStarget1} and
\ref{fig:thermoIREStarget2}, the {\em inactive}
target structure $S_1$ at $T_1 = 30^{\circ}$C was chosen to destroy
domain 5 stem-loop and unpaired PPT region, while target {\em active}
structure $S_2$ at $T_2 = 42^{\circ}$C is the experimentally determined
structure of wild-type domain 5 
FMDV IRES element. Sequence constraints were 
chosen in accordance with conservation
observed in a multiple alignment of 183 IRES elements
\cite{Fernandez11a}, where we added AUG start codon at positions 47-49
(corresponding to IRES positions 463-465).
\begin{tiny}
\begin{verbatim}
Inactive S1: ..............((((((..((((((....))))))..))))))...
  Active S2: ..(((((.(((.....)))))))).........................
Constraints: NUAGGNGACCGNAGGNCGGCNCNUUYYYYYYRNNNNNNNNNNNNNNAUG
\end{verbatim}
\end{tiny}
Using {\tt RNAiFold2T}, 24,410 solutions were generated, although an 
additional 45,442 sequences were generates using variants of target $S_1$.

\item
{\tt RNAiFold2T} solutions were discarded if any of the following criteria 
were not met:
\begin{itemize}
\item
Wild-type structures for domain 4 and domain 5 
appear as stable substructures using 
{\tt Vienna Package RNALfold -L 110 -T 42}.
\item
Domain 4 appears as a stable substructure using 
{\tt RNALfold -L 110 -T 30}
\item
Probability $Pr(S_2,T_2)$ of active conformation at $T_2=42^{\circ}$C 
exceeds $0.2$
\item
Probability $Pr(S_1,T_1)$ of inactive conformation at $T_1 = 30^{\circ}$C 
exceeds $0.2$
\item
Probability of intended target structure at intended temperature is more
than double that of unintended target, i.e.
$Pr(S_1,42)$ $/$ $Pr(S_2,42)$ $<$ $0.5$ and  
$Pr(S_2,30)/Pr(S_1,30) < 0.5$. 
\end{itemize}

\item 
Retained solutions were further filtered using various measures.
For instance, candidate 1 (Seq1) had the highest value
of $A+B$, where $A$ is
$a \cdot \left( b \cdot Pr(S_1,30) + 
c \cdot (Pr(S_1,30) - Pr(S_2,30)) \right)$ and
$B$ is
$d$ $\cdot$ $(1-$ $Pr(S_1,30))$ $+$
$e$ $\cdot$ $\left( b \right.$ $\cdot$ $Pr(S_2,42)$ $+$ 
$c \cdot$ $(Pr(S_2,42)$ $-$ $\left. Pr(S_1,42)) \right)$, and
$a=4, b=0.5, c=0.5, d=2, e=1$.
This measure was designed to select sequences where the probability of 
the intended target at the intended temperature is high, while 
probability of the unintended target is low. The measure is weighted 
to increase the likelihood of not having the inactive conformation at 
$42^{\circ}$C. In contrast,
Candidate 2 (Seq2) is one of two sequences satisfying $Pr(S_2,42)>0.3$
and $P(S_1,30) > 0.3$. See Supplementary Information (SI) for a spread sheet
of measures and their values.
\end{enumerate}
Seq1 and Seq2  consist of the following 46 nt:
Seq1 is AUAGGUGACC GGAGGGCGGC ACCUUUUUUC CAGAAAAGUA GUCGUC
(15/46 positions differ from wild-type)
and Seq2 is
GUAGGUGACC GGAGGACGGC ACCUUUUUUC CAGAAAAGUA GUCGUC
(16/46 positions differ from wild-type).
Figure~\ref{fig:experimentalResults} shows that Seq1 and Seq2
displayed an increase of approximately
50\% normalized IRES-dependent translation efficiency in RRL at 
42$^{\circ}$C versus 30$^{\circ}$C. Seq1 and Seq2 IRES elements 
displayed about 20\% lower normalized activity than the wild type 
IRES. Nonetheless, the wild type IRES was equally active at all 
temperatures tested (30, 37 and 42 $^{\circ}$C).

\subsection{Comparison of {\tt RNAiFold2T} with other software}

Here we compare the Constraint Programming (CP) and Large Neighborhood
Search (LNS) programs of {\tt RNAiFold2T} with the adaptive walk program
\SD ({\tt SD}) \cite{flamm:RNAswitch} and the genetic algorithm
\FRNA ({\tt FRNA}) \cite{frnakenstein}.
Below, we describe benchmarking results for datasets of thermosensor
target structures $S_1$ resp. $S_2$ at temperatures $T_1$ resp. $T_2$,
constructed as described in Methods.
All benchmarking was carried out on a Core2Duo PC (2.8 GHz; 2 Gbyte
memory; CentOS 5.5).

\begin{figure}[htbp]
\centering
\includegraphics[width=0.8\textwidth]{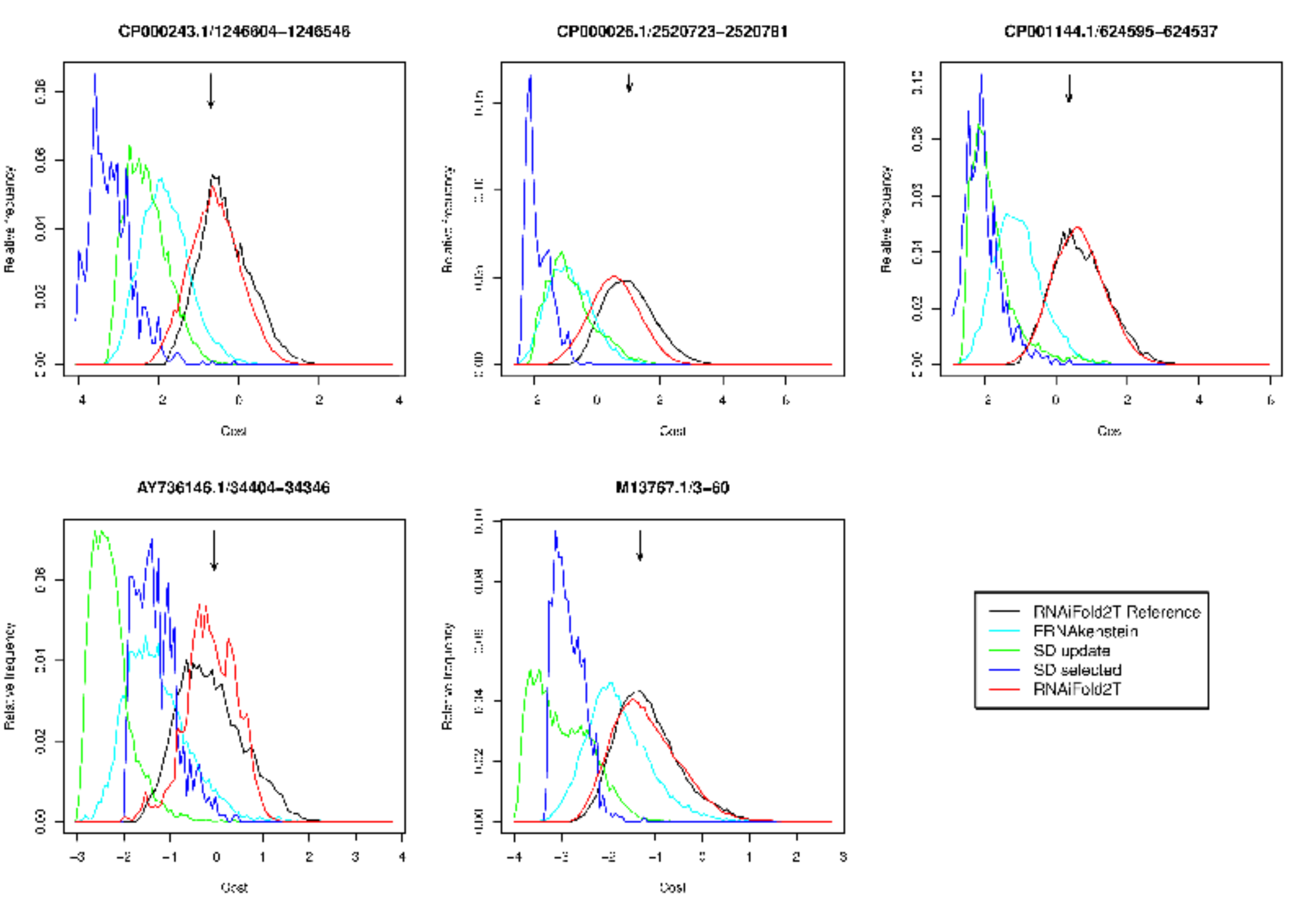}
\caption{\small Relative frequency of the cost function optimized by
\SD, for solutions returned by
{\tt RNAiFold2T}, \SD and \FRNA, given target structure
$S_1$ [resp. $S_2$] at temperature $T_1$ [resp. $T_2$] for
$\lambda$ phage CIII thermoregulators from Rfam family RF01804. 
This figure is a more generous representation of the data from
\SD and \FRNA, since all
single point mutant solutions have been added to the raw output.
(SI Figure~1 presents histograms for the raw output of 
these programs.
The {\em reference} distribution for {\em RNAiFold2T Reference}
(black curve), was produced by
running {\tt RNAiFold2T} for several days. Remaining curves are for
\FRNA (light green), \SD (dark green and purple) and {\em RNAiFold2T} (red).
Arrows indicate cost values for the real $\lambda$ phage
CIII thermoregulators from Rfam RF01804. Distribution for {\tt SD} and 
{\tt FRNA} without additional single point mutants shown in SI.
SI Figures 1,3 show clearly that cost function values for
Rfam sequences approximately equal the reference distribution mean. 
}
\label{fig:distributionsB}
\end{figure}

{\tt SD}, the first algorithm capable of designing thermoswitches, achieves 
this by optimizing the cost function given in equation (7) of
\cite{flamm:RNAswitch} -- see also equation (1) in SI.
In this context, we wanted to ascertain whether natural thermoswitches 
are optimized for this cost.  Surprisingly,
Figure~\ref{fig:distributionsB}
and Supplementary Figure~1 show that
natural thermosensor sequences from Rfam appear {\em not} to be 
optimized for the cost function used in {\tt SD}. In particular, 
{\tt SD} and {\tt FRNA} return solutions having substantially lower
cost values (i.e. more optimal) than those of natural thermosensors, 
whose cost value appears to be the mean value returned by {\tt RNAiFold2T}. 
Therefore, our benchmarking comparison of {\tt SD},
\FRNA, and {\tt RNAiFold2T} compares the number of problems solved
by each algorithm within the same amount of time.

Supplementary Information (SI) Table 1 
describes the value ordering heuristics used in
{\tt RNAiFold2T}; SI Table 2 benchmarks {\tt RNAiFold2T} 
with respect to different helix ordering and
value heuristics, using a cutoff time of 10 minutes. 
Since the data clearly demonstrates the superiority of
$overlap_2$ helix ordering, this is taken as the default for all other
benchmarks and for the web server. 
SI Table 3 [resp. SI Table 4] presents benchmarking data
for Large Neighborhood Search (LNS) from {\tt RNAiFold2T} and \SD and
\FRNA, each with a cutoff time of 30 minutes,
using Rfam thermosensor target structures of length less than 130 nt
[resp. greater than 130 nt] -- see Methods for construction of target
structures and temperatures. {\tt RNAiFold2T} has essentially the
same performance as {\tt SD} and {\tt FRNA} for shorter sequences,
while {\tt SD} performs better than other methods for longer sequences.
Target structures $S_1,S_2$ and temperatures
$T_1,T_2$ for this test are given in
SI Table 5 displays the number of 
solutions for $\lambda$ phage CIII thermoswitches
from Rfam family RF01804. Constraint Programming (CP) from
{\tt RNAiFold2T}, adaptive walk \SD  \cite{flamm:RNAswitch} and 
genetic algorithm \FRNA \cite{frnakenstein}
were run on each thermosensor for 24 hours,
forcing a restart if no new solution was found within 1 hour. Since
both \SD and \FRNA return only a single solution, for this test
we modified each as
described in Methods, resulting in two versions of \SD and one of \FRNA,
each of which returns two orders of magnitude less solutions 
than {\tt RNAiFold2T}. 
SI Table 6 lists the target structures $S_1,S_2$ and
temperatures $T_1,T_2$ of RNA thermometers of length less than 130 nt
used in the benchmarking tests. The structures for thermometers of
length 130-447 nt are too large for display, hence are available in an
Excel file in Supplementary Information.

\section{Discussion}

In this paper, we introduce the software {\tt RNAiFold2T}, capable of
solving the inverse folding problem for two or more temperatures, i.e.
generating one or more RNA sequences whose minimum free energy (MFE)
secondary structures at temperatures $T_1$ and $T_2$ [resp. $T_1,\ldots,T_m$]
are user-specified target structures $S_1$ and $S_2$ [resp. $S_1,\ldots,S_m$],
or which reports that no such solution exists. {\tt RNAiFold2T} is unique
in that it implements two different algorithms -- {\em Constraint Programming}
(CP) and {\em Large Neighborhood Search} (LNS). CP is an exact, non-heuristic
method that uses an exhaustive yet efficient branch-and-prune process, and
is the only currently available software capable of generating all solutions
or determining that no solution exists (since there are possibly exponentially
many solutions, a complete solution is feasible only for structures of modest
size). 
CP differs from what one might call a `brute-force' approach only
in that it relies on a highly efficient branch-and-prune search engine, that
propagates the effects of currently instantiated variables held within a
{\em constraint store}.
LNS uses a local search heuristic, complemented with local calls of
constraint programming to explore solutions of substructures of the target
structures. 


There exists an RNA sequence compatible with any two given secondary
more than two structures \cite{reidysStadlerSchuster} -- nevertheless,
{\tt RNAiFold2T} solves inverse folding problem for more than two temperatures.
%
\RNAiTFold is modular software, with a clear separation between 
search procedure and constraint descriptions, thus permitting the
future addition of sequence and structural constraints. In its current form,
{\tt RNAiFold2T} includes constraints for
{\em full} and/or {\em partial}
target structures or hybridization complexes at two temperatures;
a plug-in to use {\tt RNAfold} or {\tt RNAstructure} for MFE structure
computation;
IUPAC nucleotide constraints, IUPAC amino acid constraints
that require all returned RNA sequences to code specified peptides in
one or more overlapping reading frames, structural compatibility and
structural incompatibility constraints, etc. These constraints support the
design of temperature-sensitive selenocysteine insertion (SECIS) elements, 
precursor microRNAs, and mRNA domains that are targeted by microRNAs, 
etc. Since Constraint Programming (CP) is not a heuristic, unlike other
methods such as adaptive walk, genetic algorithm, etc., 
{\tt RNAiFold2T} can in principle return all 2-temperature 
inverse folding solutions, or prove that none exist.
The Large Neighborhood Search (LNS) algorithm of {\tt RNAiFold2T}
returns a single solution with approximately the same performance as
state-of-the-art approaches {\tt SD} and {\tt FRNA}.

As with the synthetic hammerhead design in \cite{Dotu.nar15},
our synthetic RNA design strategy consists of generating many solutions, 
which are prioritized for experimental validation by applying various
computational filters. In our opinion, this strategy presents advantages
over methods using {\tt SD} or \NUPACK, each of which returns
a relatively small number of sequences that are optimized with respect to
a single criterion -- in the case of {\tt SD}, this is the cost function
\cite{flamm:RNAswitch}, and in the case of \NUPACK, this is
ensemble defect \cite{Zadeh.jcc11}.

In order to ascertain the viability of our approach of not committing
to a particular cost function, we used the capabilities of \RNAiTFold
to find hundreds of thousands 
of solutions to the 2-temperature inverse folding problem
for target structures from Rfam family RF01804 
($\lambda$ phage CIII thermoregulators). 
Figure~\ref{fig:distributionsB} and SI Figure~1 show that
that the cost function value of (real) Rfam sequences
is not close to the minimum, but rather close to the average of the
distribution.  Other figures can be found in Supplementary Information,
where we investigated a variant of the cost function defined using ensemble
defect. So, although {\tt SD} benchmarking  results in Supplementary
Information indicate that cost function minimization is a good strategy to
find sequences whose MFE structures at temperatures $T_1$ resp. $T_2$ are the 
target structures $S_1$ resp. $S_2$, it appears that naturally occurring
RNA thermoswitches are not optimized for the {\tt SD} cost
function. This observation may be important for the future design of
functional synthetic thermoregulators.

\begin{figure}[htbp]
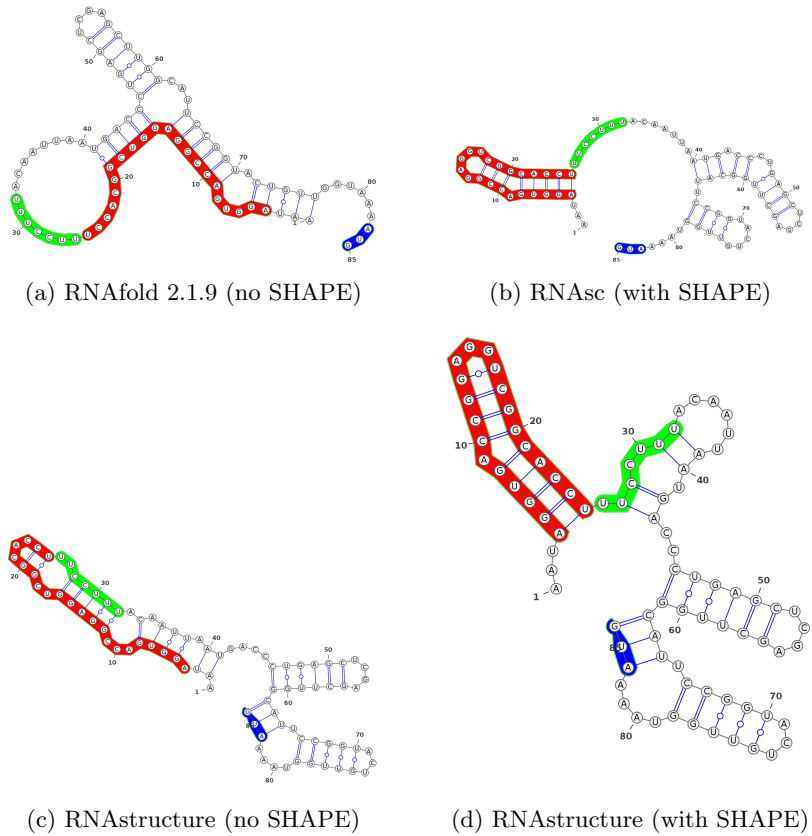

\centering
\begin{subfigure}[b]{0.45\textwidth}
\includegraphics[width=\textwidth]{FIGURES/fig5A.pdf}
\caption{\small RNAfold 2.1.9 (no SHAPE)}
\label{fig:SHAPEa}
\end{subfigure}%
\quad
\begin{subfigure}[b]{0.45\textwidth}
\includegraphics[width=\textwidth]{FIGURES/fig5B.pdf}
\caption{\small RNAsc (with SHAPE)}
\label{fig:SHAPEb}
\end{subfigure}
\\
\begin{subfigure}[b]{0.45\textwidth}
\includegraphics[width=\textwidth]{FIGURES/fig5C.pdf}
\caption{\small RNAstructure (no SHAPE)}
\label{fig:SHAPEc}
\end{subfigure}%
\quad
\begin{subfigure}[b]{0.45\textwidth}
\includegraphics[width=\textwidth]{FIGURES/fig5D.pdf}
\caption{\small RNAstructure (with SHAPE)}
\label{fig:SHAPEd}
\end{subfigure}
\caption{\small Secondary structure predictions of domain 5 of wild-type
FMDV IRES, including the first functional AUG start coden,
both with and without integration of probing data using
selective $2'$-hydroxyl acylation analyzed by primer extension (SHAPE).
%
%
(a) {\tt RNAfold 2.1.9}, which does not incorporate SHAPE data.
(b) {\tt RNAsc}
\cite{Zarringhalam.po12}, which penalizes deviations from SHAPE
data for both base paired and loop regions.
(c) {\tt RNAstructure} \cite{Reuter.bb10} without SHAPE data.
(d) {\tt RNAstructure}
\cite{Deigan.pnas09}, which penalizes deviations from SHAPE
data only for base paired regions. The polypyrimidine tract (PPT)
is unbound in only structures (a) from
{\tt RNAfold 2.1.9} and (b) from {\tt RNAsc}, compatible with the full
IRES structure appearing in Figure 1 of \cite{Fernandez11a}.
}
\label{fig:SHAPE}
\end{figure}

In designing thermo-IRES elements, we solved the 2-temperature inverse folding
problem depicted in Figure~\ref{fig:thermoIREStarget}, where the AUG start
codon was appended
at the $3'$ end of the 46 IUPAC constraint mask
NUAGGNGACC GNAGGNCGGC NCNUUYYYYY YRNNNNNNNN NNNNNN. This was
done since we assumed that the AUG start codon was located at 
position 463, as it occurs in the viral RNA.  However, experimental validation
was performed by using a construct containing the 35 nt spacer 
GAGCUCGAGC UUGGCAUUCC GGUACUGUUG GUAAA at the $3'$ end of the designed 46 nt
sequences, followed by the luciferase AUG start codon. 
It is possible that the secondary structure of the spacer might have
rendered the polypyridine tract less accessible, thus explaining the
low efficiency of thermo-IRES constructs Seq1 and Seq2.
Another issue is that IRES elements are  complex molecules, 
both in sequence and structure, which respond to more host factors 
than previously reported RNA thermometers \cite{Lozano.cov15}.
Secondary structure predictions could differ from the physical structure
due to unmodeled protein-RNA interactions, possibly unreliable free energy 
parameters at temperatures different than 37$^{\circ}$C, and due to 
variations in structure prediction software as shown in 
Figure~\ref{fig:SHAPE}. Panels (a) resp. (b) of that figure
show that the polypyrimidine tract (PPT) is unpaired in the
secondary structure of domain 5 of wild-type IRES at 37$^{\circ}$C
(hence consistent with experimental data), when
determined by {\tt RNAfold} (without SHAPE) resp. {\tt RNAsc} (with SHAPE)
\cite{Zarringhalam.po12}, while panels (c) and (d) show
the PPT is paired in the structure determined by {\tt RNAstructure}, 
both with and without SHAPE \cite{Deigan.pnas09}
(hence inconsistent with experimental data).
Due to this variation in MFE structure prediction, it may be advisable in
future synthetic RNA design projects to run {\tt RNAiFold2T} using plug-ins
for both {\tt RNAfold} and {\tt RNAstructure}, and to select sequences
which have the same desired target structures as predicted by both algorithms.
Another issue is that for many RNA thermometers, the minimum free energy (MFE)
structures at low temperature $T_1$ and high temperature $T_2$ are almost identical, 
where $T_1,T_2$ are the temperatures for which a conformational change is
reported in the literature. For instance, the MFE structures for the ROSE element
CP000009.1/1450710-1450627 at $0^{\circ}$C and $60^{\circ}$C are nearly identical.
This could be an important issue for certain software \cite{Chursov.nar13,Churkin.po14}
that predict RNA thermometers, but is of less importance in the synthetic design of
thermometers, where it is possible to exaggerate the temperature difference 
$T_2-T_1$. Perhaps future experimental work will provide more reliable energy parameters
at temperatures different than $37^{\circ}$C, an issue that affects both RNA and
DNA melting temperature predictions \cite{Rouzina.bj99}.

\section{Conclusions}
We present the software {\tt RNAiFold2T} for the multi-temperature
inverse folding problem, used to design functional thermoswitches.
The CP variant of \RNAiTFold returns two orders of magnitude more
solutions than other software, while the LNS variant,
which returns a single solution, exhibits comparable performance 
with that of existent methods.
The software design of {\tt RNAiFold2T} currently supports 
a much greater variety of user-defined structural and sequence
constraints than other methods, and moreover
can be extended to support future constraints.

Naturally occurring RNA thermometers in heat-shock and virulence genes, 
as well as all previously known instances of rationally designed thermometers
control translation initiation
by sequestering the Shine-Dalgarno (SD) ribosomal binding sequence at low
temperatures, whereby the SD becomes accessible at high temperatures due to the
melting of a hairpin. In contrast, we have used {\tt RNAiFold2T} to
design a temperature-regulated
internal ribosomal entry site (IRES) element by ensuring the presence at 
high temperatures of the domain 5 stem-loop and downstream single-stranded 
pyrimidine (Py) tract, both
located upstream of the functional initiation codon and both
known to be important for IRES functionality \cite{Lozano.cov15}. 
At low temperatures, our thermo-IRES element is designed to adopt a 
conformation that down-regulates protein product by disrupting both the 
domain 5 stem-loop and sequestering the Py tract.
We showed that our rationally designed thermo-IRES elements are functional,
where the cap-independent translational efficiency is approximately
50\% higher at 42$^{\circ}$C than at 30$^{\circ}$C; however,  since
the focus of this paper is primarily to describe a new method, we have not
taken steps to improve efficiency using error-prone mutagenesis and selection.


\section{Funding}

Research of the Clote Lab was supported
by National Science Foundation grant
DBI-1262439.  Any opinions, findings,
and conclusions or recommendations expressed in this material are
those of the authors and do not necessarily reflect the views of the
National Science Foundation.
Research of the Martinez-Salas Lab was 
supported by the Spanish Ministry of Economy and Competitiveness 
(MINECO) [CSD2009-00080, BFU2011-25437, BFU2014-54564] and by an 
Institutional Grant from Fundaci{\'o}n Ram{\'o}n Areces.

\bibliographystyle{plain}

\begin{thebibliography}{10}

\bibitem{Altuvia.jmb89}
S.~Altuvia, D.~Kornitzer, D.~Teff, and A.~B. Oppenheim.
\newblock Alternative m{RNA} structures of the c{III} gene of bacteriophage
  lambda determine the rate of its translation initiation.
\newblock {\em J. Mol. Biol.}, 210(2):265--280, November 1989.

\bibitem{Andronescu.b07}
M.~Andronescu, A.~Condon, H.~H. Hoos, D.~H. Mathews, and K.~P. Murphy.
\newblock Efficient parameter estimation for {RNA} secondary structure
  prediction.
\newblock {\em Bioinformatics}, 23(13):i19--i28, July 2007.

\bibitem{Bae.pnas00}
W.~Bae, B.~Xia, M.~Inouye, and K.~Severinov.
\newblock Escherichia coli {CspA}-family {RNA} chaperones are transcription
  antiterminators.
\newblock {\em Proc. Natl. Acad. Sci. U.S.A.}, 97(14):7784--7789, July 2000.

\bibitem{ROSEswitch}
S.~Chowdhury, C.~Ragaz, E.~Kreuger, and F.~Narberhaus.
\newblock Temperature-controlled structural alterations of an {RNA}
  thermometer.
\newblock {\em J. Biol. Chem.}, 278(48):47915--47921, November 2003.

\bibitem{Churkin.po14}
A.~Churkin, A.~Avihoo, M.~Shapira, and D.~Barash.
\newblock {RNAthermsw}: direct temperature simulations for predicting the
  location of {RNA} thermometers.
\newblock {\em PLoS. One.}, 9(4):e94340, 2014.

\bibitem{Chursov.nar13}
A.~Chursov, S.~J. Kopetzky, G.~Bocharov, D.~Frishman, and A.~Shneider.
\newblock {RNAtips}: {Analysis} of temperature-induced changes of {RNA}
  secondary structure.
\newblock {\em Nucleic. Acids. Res.}, 41(Web):W486--W491, July 2013.

\bibitem{Deigan.pnas09}
K.~E. Deigan, T.~W. Li, D.~H. Mathews, and K.~M. Weeks.
\newblock Accurate {SHAPE}-directed {RNA} structure determination.
\newblock {\em Proc. Natl. Acad. Sci. U.S.A.}, 106(1):97--102, January 2009.

\bibitem{Dotu.nar15}
I.~Dotu, J.~A. Garcia-Martin, B.~L. Slinger, V.~Mechery, M.~M. Meyer, and
  P.~Clote.
\newblock Complete {RNA} inverse folding: computational design of functional
  hammerhead ribozymes.
\newblock {\em Nucleic. Acids. Res.}, 42(18):11752--11762, February 2015.

\bibitem{Dotu.rb13}
I.~Dotu, G.~Lozano, P.~Clote, and E.~Martinez-Salas.
\newblock Using {RNA} inverse folding to identify {IRES}-like structural
  subdomains.
\newblock {\em RNA. Biol.}, 10(12):1842--1852, December 2013.

\bibitem{Fernandez11a}
N.~Fernandez, O.~Fernandez-Miragall, J.~Ramajo, A.~García-Sacristan,
  N.~Bellora, E.~Eyras, C.~Briones, and E.~Martinez-Salas.
\newblock Structural basis for the biological relevance of the invariant apical
  stem in {IRES}-mediated translation.
\newblock {\em Nucleic Acids Res.}, 39:8572--8585, 2011.

\bibitem{FernandezChamorro.nar14}
J.~Fernandez-Chamorro, D.~Pineiro, J.~M. Gordon, J.~Ramajo,
  R.~Francisco-Velilla, M.~J. Macias, and E.~Martinez-Salas.
\newblock Identification of novel non-canonical {RNA}-binding sites in {Gemin5}
  involved in internal initiation of translation.
\newblock {\em Nucleic. Acids. Res.}, 42(9):5742--5754, May 2014.

\bibitem{flamm:RNAswitch}
C.~Flamm, I.L. Hofacker, S.~Mauer-Stroh, P.F. Stadler, and M.~Zehl.
\newblock Design of multi-stable {RNA} molecules.
\newblock {\em RNA}, 7:254--265, 2001.

\bibitem{GarciaMartin.nar15}
J.~A. Garcia-Martin, I.~Dotu, and P.~Clote.
\newblock {RNAiFold} 2.0: a web server and software to design custom and
  {Rfam}-based {RNA} molecules.
\newblock {\em Nucleic. Acids. Res.}, 43(W1):W513--W521, July 2015.

\bibitem{GarciaMartin12}
J.A. Garcia-Martin, P.~Clote, and I.~Dotu.
\newblock {RNAiFold}: {A} constraint programming algorithm for {RNA} inverse
  folding and molecular design.
\newblock {\em Journal of Bioinformatics and Computational Biology},
  11(2):1350001, 2012.

\bibitem{HoynesOConnor.nar15}
A.~Hoynes-O'Connor, K.~Hinman, L.~Kirchner, and T.~S. Moon.
\newblock De novo design of heat-repressible {RNA} thermosensors in {E}. coli.
\newblock {\em Nucleic. Acids. Res.}, 43(12):6166--6179, July 2015.

\bibitem{Johansson.c02}
J.~Johansson, P.~Mandin, A.~Renzoni, C.~Chiaruttini, M.~Springer, and
  P.~Cossart.
\newblock An {RNA} thermosensor controls expression of virulence genes in
  {Listeria} monocytogenes.
\newblock {\em Cell}, 110(5):551--561, September 2002.

\bibitem{Kortmann.nrm12}
J.~Kortmann and F.~Narberhaus.
\newblock Bacterial {RNA} thermometers: molecular zippers and switches.
\newblock {\em Nat. Rev. Microbiol.}, 10(4):255--265, April 2012.

\bibitem{Kortmann.nar11}
J.~Kortmann, S.~Sczodrok, J.~Rinnenthal, H.~Schwalbe, and F.~Narberhaus.
\newblock Translation on demand by a simple {RNA}-based thermosensor.
\newblock {\em Nucleic. Acids. Res.}, 39(7):2855--2868, April 2011.

\bibitem{Kuhn.jv90}
R.~Kuhn, N.~Luz, and E.~Beck.
\newblock Functional analysis of the internal translation initiation site of
  foot-and-mouth disease virus.
\newblock {\em J. Virol.}, 64(10):4625--4631, October 1990.

\bibitem{nanoscaleThermometer}
J.~Lee and J.A. Kotov.
\newblock Thermometer design at the nanoscale.
\newblock {\em Nano Today}, 2(1):48--51, 2007.

\bibitem{Lorenz.amb11}
R.~Lorenz, S.~H. Bernhart, C.~H{\"o}ner~zu Siederdissen, H.~Tafer, C.~Flamm,
  P.~F. Stadler, and I.~L. Hofacker.
\newblock Viennarna {Package} 2.0.
\newblock {\em Algorithms. Mol. Biol.}, 6:26, 2011.

\bibitem{Lozano.fj14}
G.~Lozano, N.~Fernandez, and E.~Martinez-Salas.
\newblock Magnesium-dependent folding of a picornavirus {IRES} element
  modulates {RNA} conformation and e{IF4G} interaction.
\newblock {\em FEBS J.}, 281(16):3685--3700, August 2014.

\bibitem{Lozano.cov15}
G.~Lozano and E.~Martinez-Salas.
\newblock Structural insights into viral {IRES}-dependent translation
  mechanisms.
\newblock {\em Curr. Opin. Virol.}, 12:113--120, June 2015.

\bibitem{Lozano.rb15}
G.~Lozano, A.~Trapote, J.~Ramajo, X.~Elduque, A.~Grandas, J.~Robles,
  E.~Pedroso, and E.~Martinez-Salas.
\newblock Local {RNA} flexibility perturbation of the {IRES} element induced by
  a novel ligand inhibits viral {RNA} translation.
\newblock {\em RNA. Biol.}, 12(5):555--568, 2015.

\bibitem{frnakenstein}
R.~B. Lyngso, J.~W. Anderson, E.~Sizikova, A.~Badugu, T.~Hyland, and J.~Hein.
\newblock Frnakenstein: multiple target inverse {RNA} folding.
\newblock {\em BMC. Bioinformatics}, 13:260, 2012.

\bibitem{Nawrocki.nar14}
E.~P. Nawrocki, S.~W. Burge, A.~Bateman, J.~Daub, R.~Y. Eberhardt, S.~R. Eddy,
  E.~W. Floden, P.~P. Gardner, T.~A. Jones, J.~Tate, and R.~D. Finn.
\newblock Rfam 12.0: updates to the {RNA} families database.
\newblock {\em Nucleic Acids Res.}, 0(O):O, November 2014.

\bibitem{Neupert.nar08}
J.~Neupert, D.~Karcher, and R.~Bock.
\newblock Design of simple synthetic {RNA} thermometers for
  temperature-controlled gene expression in {Escherichia} coli.
\newblock {\em Nucleic. Acids. Res.}, 36(19):e124, November 2008.

\bibitem{Nocker.nar01}
A.~Nocker, T.~Hausherr, S.~Balsiger, N.~P. Krstulovic, H.~Hennecke, and
  F.~Narberhaus.
\newblock A m{RNA}-based thermosensor controls expression of rhizobial heat
  shock genes.
\newblock {\em Nucleic. Acids. Res.}, 29(23):4800--4807, December 2001.

\bibitem{Ogawa.r11}
A.~Ogawa.
\newblock Rational design of artificial riboswitches based on ligand-dependent
  modulation of internal ribosome entry in wheat germ extract and their
  applications as label-free biosensors.
\newblock {\em RNA.}, 17(3):478--488, March 2011.

\bibitem{Pineiro.nar13}
D.~Pineiro, N.~Fernandez, J.~Ramajo, and E.~Martinez-Salas.
\newblock Gemin5 promotes {IRES} interaction and translation control through
  its {C}-terminal region.
\newblock {\em Nucleic. Acids. Res.}, 41(2):1017--1028, January 2013.

\bibitem{reidysStadlerSchuster}
C.~Reidys, P.F. Stadler, and P.~Schuster.
\newblock Generic properties of combinatory maps: neutral networks of {RNA}
  secondary structures.
\newblock {\em Bull Math Biol.}, 59(2):339--397, 1997.

\bibitem{Reuter.bb10}
J.~S. Reuter and D.~H. Mathews.
\newblock {RNAstructure}: software for {RNA} secondary structure prediction and
  analysis.
\newblock {\em BMC. Bioinformatics}, 11:129, 2010.

\bibitem{Rinnenthal.nar11}
J.~Rinnenthal, B.~Klinkert, F.~Narberhaus, and H.~Schwalbe.
\newblock Modulation of the stability of the {Salmonella} four{U}-type {RNA}
  thermometer.
\newblock {\em Nucleic. Acids. Res.}, 39(18):8258--8270, October 2011.

\bibitem{Rouzina.bj99}
I.~Rouzina and V.~A. Bloomfield.
\newblock Heat capacity effects on the melting of {DNA}. 1. {General} aspects.
\newblock {\em Biophys. J.}, 77(6):3242--3251, December 1999.

\bibitem{Saragliadis.rb13}
A.~Saragliadis, S.~S. Krajewski, C.~Rehm, F.~Narberhaus, and J.~S. Hartig.
\newblock Thermozymes: {Synthetic} {RNA} thermometers based on ribozyme
  activity.
\newblock {\em RNA. Biol.}, 10(6):1010--1016, June 2013.

\bibitem{Torok.pnas01}
Z.~Torok, P.~Goloubinoff, I.~Horvath, N.~M. Tsvetkova, A.~Glatz, G.~Balogh,
  V.~Varvasovszki, D.~A. Los, E.~Vierling, J.~H. Crowe, and L.~Vigh.
\newblock Synechocystis {HSP17} is an amphitropic protein that stabilizes
  heat-stressed membranes and binds denatured proteins for subsequent
  chaperone-mediated refolding.
\newblock {\em Proc. Natl. Acad. Sci. U.S.A.}, 98(6):3098--3103, March 2001.

\bibitem{Turner.nar10}
D.~H. Turner and D.~H. Mathews.
\newblock {NNDB}: the nearest neighbor parameter database for predicting
  stability of nucleic acid secondary structure.
\newblock {\em Nucleic. Acids. Res.}, 38(Database):D280--D282, January 2010.

\bibitem{Wachsmuth.nar13}
M.~Wachsmuth, S.~Findeiss, N.~Weissheimer, P.~F. Stadler, and M.~Morl.
\newblock De novo design of a synthetic riboswitch that regulates transcription
  termination.
\newblock {\em Nucleic. Acids. Res.}, 41(4):2541--2551, February 2013.

\bibitem{Waldminghaus.mgg07}
T.~Waldminghaus, L.~C. Gaubig, and F.~Narberhaus.
\newblock Genome-wide bioinformatic prediction and experimental evaluation of
  potential {RNA} thermometers.
\newblock {\em Mol. Genet. Genomics.}, 278(5):555--564, November 2007.

\bibitem{Waldminghaus.mm07}
T.~Waldminghaus, N.~Heidrich, S.~Brantl, and F.~Narberhaus.
\newblock Fouru: a novel type of {RNA} thermometer in {Salmonella}.
\newblock {\em Mol. Microbiol.}, 65(2):413--424, July 2007.

\bibitem{Waldminghaus.bc08}
T.~Waldminghaus, J.~Kortmann, S.~Gesing, and F.~Narberhaus.
\newblock Generation of synthetic {RNA}-based thermosensors.
\newblock {\em Biol. Chem.}, 389(10):1319--1326, October 2008.

\bibitem{Wieland.cb07}
M.~Wieland and J.~S. Hartig.
\newblock {RNA} quadruplex-based modulation of gene expression.
\newblock {\em Chem. Biol.}, 14(7):757--763, July 2007.

\bibitem{Zadeh.jcc11}
J.~N. Zadeh, B.~R. Wolfe, and N.~A. Pierce.
\newblock Nucleic acid sequence design via efficient ensemble defect
  optimization.
\newblock {\em J. Comput. Chem.}, 32(3):439--452, February 2011.

\bibitem{Zarringhalam.po12}
K.~Zarringhalam, M.~M. Meyer, I.~Dotu, J.~H. Chuang, and P.~Clote.
\newblock Integrating chemical footprinting data into {RNA} secondary structure
  prediction.
\newblock {\em PLoS. One.}, 7(10):e45160, 2012.

\bibitem{zuker:mfoldWebserver}
M.~Zuker.
\newblock Mfold web server for nucleic acid folding and hybridization
  prediction.
\newblock {\em Nucleic Acids Res.}, 31(13):3406--3415, 2003.

\end{thebibliography}

\begin{footnotesize}

\end{footnotesize}

\end{document}